\documentclass[12pt]{article}
\usepackage{epsf}
\def\be{\begin{equation}}
\def\ee{\end{equation}}
\def\bea{\begin{eqnarray}}
\def\eea{\end{eqnarray}}

\begin{document}
\begin{titlepage}
\begin{center}
{\Large \bf William I. Fine Theoretical Physics Institute \\
University of Minnesota \\}
\end{center}
\vspace{0.2in}
\begin{flushright}
FTPI-MINN-18/19 \\
UMN-TH-3802/18 \\
October 2018 \\
\end{flushright}
\vspace{0.3in}
\begin{center}
{\Large \bf $Z_c(4100)$ and $Z_c(4200)$ as hadrocharmonium.
\\}
\vspace{0.2in}
{\bf  M.B. Voloshin  \\ }
William I. Fine Theoretical Physics Institute, University of
Minnesota,\\ Minneapolis, MN 55455, USA \\
School of Physics and Astronomy, University of Minnesota, Minneapolis, MN 55455, USA \\ and \\
Institute of Theoretical and Experimental Physics, Moscow, 117218, Russia
\\[0.2in]

\end{center}

\vspace{0.2in}

\begin{abstract}
It is suggested that the recently reported charged charmoniumlike resonance $Z_c(4100)$ and the previously observed $Z_c(4200)$ are two states of hadrocharmonium, related by the charm quark spin symmetry in the same way as the lowest charmonium states $\eta_c$ and $J/\psi$. Namely, the $Z_c(4100)$ is (dominantly) the $\eta_c$ embedded in an light-quark excitation with quantum numbers of a pion, while the $Z_c(4200)$ is a similar four quark state containing $J/\psi$ instead of $\eta_c$. It is argued that this picture suggests certain relations between the properties of the two exotic resonances and a distinctive pattern of their decays that can be tested experimentally.  It is also pointed out that due to the presence of a spatially compact $c \bar c$ charmonium the discussed exotic $Z_c$ states may be produced as resonances in the $s$ channel in $\bar p p$ annihilation.
  \end{abstract}
\end{titlepage}
The charged charmoniumlike resonance $Z_c(4100)$ most recently reported~\cite{lhcbzc} in the channel $\eta_c \pi$ in the decays $B^0 \to Z_c(4100)^- K^+ \to \eta_c \pi^- K^+$ is the latest addition to the growing set of intriguing exotic states observed at the onset of the charm-anticharm and bottom-antibottom open flavor thresholds. These exotic states necessarily contain a light quark and antiquark pair in addition to a heavy quark and antiquark, e.g. the quark  content of the $Z_c(4100)^-$ has to be $c \bar c d \bar u$. The existing theoretical approaches to description of such four quark states concentrate on finding some simplification in the cases where the multiquark dynamics can be approximated by dominant two-body correlations. (Recent reviews of the experimental data and of the theoretical approaches can be found in Refs.~\cite{dsz,Guo17} and \cite{Ali17}.) It is quite natural to expect that  such dominant correlations are likely to be different in different resonances. In particular, some exotic resonances that are very close in mass to a threshold for a heavy meson and antimeson can be considered as `molecules'~\cite{vo}, i.e. as dominated by the mesons moving at relatively large distances. This picture is likely applicable to the $Z_b(10610)$ and $Z_b(10650)$ bottomoniumlike resonances~\cite{bellez} and to the $Z_c(3900)$~\cite{besz39} and $Z_c(4020)$\cite{besz40} in the charmoniumlike sector. Other exotic states with mass not being particularly close to any heavy meson pair threshold may fall into a somewhat different category. Namely, they may have the structure of `hadroquarkonium'~\cite{mvch,dv}, i.e. of a specific state of a heavy quarkonium bound, by a gluon analog of van der Waals interaction, inside an excited state of light-quark matter. The embedded heavy quarkonium state is only slightly deformed by the coupling to the light matter and dominates in the decays of the hadroquarkonium, where the light hadronic excitation produces one or more light hadrons.  Such structure would then explain the observed property of some resonances to dominantly decay into a specific state of heavy quarkonium and not the others, e.g. $Z_c(4430) \to \psi(2S) \pi$~\cite{pdg}, as well as an apparent suppression of the decay of such resonances into heavy meson pairs with open flavor\cite{dgv}. The purpose of the present paper is to point out that the most recently found resonance $Z_c(4100)$ and the previously observed one $Z_c(4200)$~\cite{bellezc} are likely two states of hadrocharmonium and are related to each other in that they have the same `hosting' excited light matter state, and the embedded charmonium states are the heavy quark spin symmetry (HQSS) partners: respectively $\eta_c$ and $J/\psi$. 

Indeed, the $Z_c(4100)$ resonance decays to $\eta_c \pi$ and its preferred by the data \cite{lhcbzc} quantum numbers are $J^P=0^+$, which are most appropriate for an $S$ wave resonance made from $\eta_c$ embedded in an excited light hadronic state $X$ with quantum numbers of a pion $J^P=0^-$. The de-excitation of the state $X$ results in emission of the pion and the $\eta_c$ becoming unbound. Within the HQSS limit the $S$ wave binding of the $J/\psi$ with the same state $X$ should then also exist and make a resonance with the quntum numbers $J^P=1^+$ and with mass heavier than $Z_c(4100)$ by the $J/\psi - \eta_c$ mass splitting, i.e. by  approximately 113\,MeV.  A brief examination of the data on $Z_c(4200)$ strongly suggests that this is indeed the HQSS partner of the $Z_c(4100)$. Namely, $J^P=1^+$ are the preferred by the data~\cite{bellezc} quantum numbers for the heavier resonance and the masses are measured as: $M[Z_c(4100)]= \left ( 4096 \pm 20 ^{+18}_{-22} \right )$\,MeV and $M[Z_c(4200)]= \left ( 4196^{+31+17}_{-29-13} \right )$\,MeV. Furthermore, the width of the dominant decays $Z_c(4100) \to \eta_c \pi$ and $Z_c(4200) \to J/\psi \pi$ should be the the same in the HQSS limit. The experimental fits for the total widths, $\Gamma[Z_c(4100)]= \left ( 152 \pm 58^{+60}_{-35} \right )$\,MeV and $\Gamma[Z_c(4200)]= \left ( 370 ^{+70 + 70}_{-70-132} \right )$\,MeV, at least do not contradict this prediction, although, admittedly, with a considerable uncertainty. In what follows it will be assumed that the spin-parity quantum numbers of the discussed resonances are the ones preferred by the experiments: $0^+$ for the $Z_c(4100)$ and $1^+$ for $Z_c(4200)$.

Moreover, the production characteristics in the $B$ meson decays for the two exotic resonances appear to be quite similar: ${\cal B}[B^0 \to Z_c(4100)^- K^+] \approx 1.9 \times 10^{-5}$ and ${\cal B}[B^0 \to Z_c(4200)^- K^+] \approx 2.2 \times 10^{-5}$. Only the central values for the data are quoted here while sizable experimental errors, arising from a number of sources, can be found in the original papers \cite{lhcbzc} and \cite{bellezc}. In this case, again modulo a considerable uncertainty, the relative yield of both resonances is in line with the relative yield of $\eta_c$ and $J/\psi$ in the decays $B \to \eta_c \pi K$ and $B \to J/\psi \pi K$. Such behavior is the one that is to be expected in the discussed hadrocharmonium picture with factorization of the production process: the spatially compact charmonium states $\eta_c$ or $J/\psi$ are produced at short distances from the $c \bar c$ pair emerging in the weak $B$ decay and then at longer distances the charmonium is captured into hadrocharmonium. It should be noted however, that generally the relative yield of $\eta_c$ and $J/\psi$ can vary over the phase space, and an accurate similarity of the relative yield of the $Z_c(4100)$ and $Z_c(4200)$ resonances to that of $\eta_c$ and $J/\psi$ should be expected only in comparison with the non-resonant background of $\eta_c \pi$ and $J/\psi \pi$  around the same invariant mass as the resonances. Indeed, in this region the rate of capture of charmonium to the bound state should not depend on the heavy quark spin (due to HQSS), so that one can estimate 
\be
{{\cal B}[B^0 \to Z_c(4100)^- K^+] \over {\cal B}[B^0 \to Z_c(4200)^- K^+]} \approx \left . {{\cal B}[B^0 \to \eta_c \pi^- K^+] \over {\cal B}[B^0 \to J/\psi \pi^- K^+]} \right |_{M(c \bar c \pi) \approx M(Z_c)}~.
\label{simy}
\ee

It can be also pointed out that in absolute terms the yield of both discussed $Z_c$ resonances in the hard processes of $B$ decays is not very small, and is certainly much larger than one would expect~\cite{bk} for a `molecular' state. This behavior indicates that the heavy $c \bar c$ pair in these resonances is at sufficiently short distances to have a significant overlap with the short-distance state of this pair produced in the weak $B$ decays. Based on this property it can be expected that the $Z_c(4100)$ and $Z_c(4200)$ resonances can be produced at a measurable rate not only in the $B$ decays, but also in high-energy proton-proton and proton-antiproton collisions, similarly to the known production of the $X(3872)$ charmoniumlike resonance~\footnote{In the $X(3872)$ a presence of a hard $c \bar c$ pair is likely due to mixing with pure $^3P_1$ charmonium (a discussion can be found  e.g. in the review \cite{mvch}) Clearly, such mixing is impossible for the charged $Z_c$ resonances, however a spatially compact $c \bar c$ pair is present in the embedded charmonium.}. The large width of the discussed $Z_c$ resonances may present a considerable challenge for separating them from background in high energy hadronic collisions, unlike in the case of very narrow $X(3872)$. This difficulty however may be somewhat relaxed if the discussed resonances are produced in a lower-energy formation type setting, e.g. their neutral isotopic partners can be produced by formation in $\bar p p$ collisions, $\bar p p \to Z_c^0$, in the future PANDA experiment~\cite{panda}, although at present it appears impossible to estimate the formation cross section.  

The similarity of the $Z_c(4100)$ and $Z_c(4200)$ resonances in the discussed hadrocharmonium picture is based on the HQSS similarity between $\eta_c$ and $J/\psi$. However effects of violation of the spin symmetry are to be expected, especially for the charmed quarks, whose mass is not very large as compared to $\Lambda_{QCD}$. A typical scale of HQSS violation in charmonium can be estimated as being in the ballpark of $O(0.1)$ (in the rate) from the observed relative rate of the HQSS violating and conserving decays: $\psi(2S) \to J/\psi \eta$ and $\psi(2S) \to J/\psi \pi \pi$. In the discussed $Z_c$ resonances an obvious effect of HQSS breaking would be a presence of sub-dominant `cross' decays: $Z_c(4100) \to J/\psi \rho$ and $Z_c(4200) \to \eta_c \rho$. Both decays are $S$ wave processes and can be induced by the M1 term in the Hamiltonian $H$ for QCD multipole expansion~\cite{gottfried,mv79,mvch}, describing the interaction of the chromomagnetic moments of the charmed quark and antiquark with the chromomagnetic gluon field $B^a$:
\be
H_{M1}= - {1 \over 2 m_c} \,\xi^a \, \left ( {\vec \Delta} \cdot {\vec B}^a \right ) ~,
\label{hm1}
\ee
where $\xi^a=t_1^a-t_2^a$ is the difference of the color generators acting on the heavy quark and the antiquark, and $\vec \Delta = \vec s_1 - \vec s_2$ is a similar difference of the spin operators. The operator $\Delta$ describes the spin-flip amplitude for transition between the $^1S_0$ ($\eta_c$) and $^3S_1$ ($J/\psi$) spin states of the $c \bar c$ pair, which amplitude obviously has the same strength in either direction. One thus readily concludes that the leading HQSS breaking term (\ref{hm1}) gives rise to $S$ wave amplitudes of the decays $Z_c(4100) \to J/\psi \rho$ and $Z_c(4200) \to \eta_c \rho$ being proportional to the same constant $C$:
\be
A[Z_c(4100) \to J/\psi \rho]= C \, ({\vec \psi} \cdot {\vec \rho})~;~~~~ A[Z_c(4200) \to \eta_c \rho]= C \, ({\vec Z} \cdot {\vec \rho})~,
\label{azcr}
\ee
where $\vec \rho$, $\vec \psi$ and $\vec Z$ are the polarization amplitudes of respectively $\rho$ $J/\psi$ and $Z_c(4200)$. Using the expressions in Eq. (\ref{azcr}) one readily gets the relation between the decay rates:
\be
\Gamma[Z_c(4100) \to J/\psi \rho] \approx 3 \, \Gamma[Z_c(4200) \to \eta_c \rho]~,
\label{rzcr}
\ee
modulo a slight difference in the phase space, whose effect in these $S$ wave processes should be rather small. Given the overall strength of HQSS breaking in charmonium, the branching fractions for these decays should be from several percent to a few tens percent. 

The dominance of a specific charmonium state in decays of hadrocharmonium is not expected to be absolute, so that other heavy states can be produced in some fraction of the decays~\cite{mvch,dv}. For the discussed $Z_c$ resonances this implies that some sub-dominant fraction of their decays should also go into channels with orbitally or radially excited charmonium. For the radial excitation one should thus expect existence of the decays $Z_c(4100) \to \eta_c(2S) \pi$ and $Z_c(4200) \to \psi(2S) \pi$. Although it is not possible to reliably estimate the absolute rates of these processes, in the HQSS limit the rates should be the same:
\be
 \Gamma[Z_c(4100) \to \eta_c(2S) \pi] \approx  \Gamma[Z_c(4200) \to \psi(2S) \pi]~.
\label{2sr}
\ee

For the decays to orbitally excited $P$-wave charmonium with emission of a single pion the parity and G-parity conservation leaves open only HQSS breaking decay channels: $Z_c(4100) \to \chi_{c1} \pi$ and $Z_c(4200) \to h_c \pi$, with both being $P$ wave processes. Using, as above, the form of leading HQSS breaking spin-flip Hamiltonian (\ref{hm1}), one readily finds that  these decays are described by the amplitude of the same strength, so that their relative rates can be estimated as
\be
{\Gamma[Z_c(4200) \to h_c \pi]   \over \Gamma[Z_c(4100) \to \chi_{c1} \pi]} \approx \left ( {p_2 \over p_1} \right )^3 \approx 1.5~,
\label{rp21}
\ee
where the kinematical difference in the phase space factor is numerically somewhat important due to the $P$-wave emission of the pion. ($p_1$ and $p_2$ are the values of the pion momentum in the two processes.) It should be also mentioned that the latter rates are likely quite small. Indeed these decays require both an excitation of the charmonium in the decay and HQSS breaking. Thus their branching fractions can be expected at the order of one percent. 

In summary. The data on the charmoniumlike exotic resonances $Z_c(4100)$ \cite{lhcbzc} and $Z_c(4200)$ \cite{bellezc} invite a suggestion that these are four quark states of the hadrocharmonium type, i.e. with respectively charmonium $\eta_c$ and $J/\psi$ embedded in an $S$ wave inside the same excited light-quark state $X$ with the quantum numbers of a pion $I^G(J^P) = 1^-(0^-)$. Such structure reproduces the preferred by the experiments quantum numbers of both resonances. The embedded charmonium states are related by HQSS, so that the spin symmetry can also be applied to the $Z_c$ resonances, implying that the difference of their mass should be close to that between $J/\psi$ and $\eta_c$ and that the widths of their dominant decays, into $\eta_c \pi$ and $J/\psi \pi$ respectively, should be the same. The existing data, still with a large uncertainty, are in agreement with these similarity relations for masses and widths. Furthermore, the hadrocharmonium picture suggests (due to HQSS) a similarity of the yield of both states in $B$ decays [Eq.(\ref{simy})], as well as the very fact of an observable yield of the discussed exotic resonances in hard processes, such as the $B$ decays and also hadronic collisions. Among the latter processes it may be possible to search for the formation of the $Z_c$ resonances in the $s$ channel in $\bar p p$ collisions of the type planned to be studied in the PANDA experiment. The HQSS and its breaking by the spin-dependent Hamiltonian (\ref{hm1}) result in relations between subdominant decays of $Z_c(4100)$ and $Z_c(4200)$ given by Eqs.~(\ref{2sr}) and (\ref{rzcr}, \ref{rp21}). Thus the discussed hadrocharmonium interpretation of the $Z_c$ resonances predicts a distinctive pattern of their dominant and subdominant properties. A test of these predictions in future studies may thus be instrumental in understanding the multiquark dynamics of exotic hadrons.


This work is supported in part by U.S. Department of Energy Grant No.\ DE-SC0011842.


\begin{thebibliography}{99}

\bibitem{lhcbzc} 
  R.~Aaij {\it et al.} [LHCb Collaboration],
  [arXiv:1809.07416 [hep-ex]].

\bibitem{dsz} 
  S.~L.~Olsen, T.~Skwarnicki and D.~Zieminska,
  Rev.\ Mod.\ Phys.\  {\bf 90}, no. 1, 015003 (2018)
  doi:10.1103/RevModPhys.90.015003
  [arXiv:1708.04012 [hep-ph]].
	
\bibitem{Guo17} 
  F.~K.~Guo, C.~Hanhart, U.~G.~Mei{\ss}ner, Q.~Wang, Q.~Zhao and B.~S.~Zou,
  Rev.\ Mod.\ Phys.\  {\bf 90}, no. 1, 015004 (2018)
  doi:10.1103/RevModPhys.90.015004
  [arXiv:1705.00141 [hep-ph]].
	
\bibitem{Ali17} 
  A.~Ali, J.~S.~Lange and S.~Stone,
  Prog.\ Part.\ Nucl.\ Phys.\  {\bf 97}, 123 (2017)
  doi:10.1016/j.ppnp.2017.08.003
  [arXiv:1706.00610 [hep-ph]].
	

\bibitem{vo} 
  M.~B.~Voloshin and L.~B.~Okun,
  JETP Lett.\  {\bf 23}, 333 (1976)
  [Pisma Zh.\ Eksp.\ Teor.\ Fiz.\  {\bf 23}, 369 (1976)].
	
\bibitem{bellez} 
  A.~Bondar {\it et al.} [Belle Collaboration],
  Phys.\ Rev.\ Lett.\  {\bf 108}, 122001 (2012)
  doi:10.1103/PhysRevLett.108.122001
  [arXiv:1110.2251 [hep-ex]].
	
\bibitem{besz39} 
  M.~Ablikim {\it et al.} [BESIII Collaboration],
  Phys.\ Rev.\ Lett.\  {\bf 110}, 252001 (2013)
  doi:10.1103/PhysRevLett.110.252001
  [arXiv:1303.5949 [hep-ex]].
	
\bibitem{besz40} 
  M.~Ablikim {\it et al.} [BESIII Collaboration],
  Phys.\ Rev.\ Lett.\  {\bf 111}, no. 24, 242001 (2013)
  doi:10.1103/PhysRevLett.111.242001
  [arXiv:1309.1896 [hep-ex]].
	
\bibitem{mvch} 
  M.~B.~Voloshin,
  Prog.\ Part.\ Nucl.\ Phys.\  {\bf 61}, 455 (2008)
  doi:10.1016/j.ppnp.2008.02.001
  [arXiv:0711.4556 [hep-ph]].
	
\bibitem{dv} 
  S.~Dubynskiy and M.~B.~Voloshin,
  Phys.\ Rev.\ D {\bf 77}, 014013 (2008)
  doi:10.1103/PhysRevD.77.014013
  [arXiv:0709.4474 [hep-ph]].
	
\bibitem{pdg}
M. Tanabashi {\it et al.} [Particle Data Group], Phys.\ Rev.\ D {\bf 98}, 030001 (2018).

\bibitem{dgv} 
  S.~Dubynskiy, A.~Gorsky and M.~B.~Voloshin,
  Phys.\ Lett.\ B {\bf 671}, 82 (2009)
  doi:10.1016/j.physletb.2008.11.040
  [arXiv:0804.2244 [hep-th]].
	
\bibitem{bellezc} 
  K.~Chilikin {\it et al.} [Belle Collaboration],
  Phys.\ Rev.\ D {\bf 90}, no. 11, 112009 (2014)
  doi:10.1103/PhysRevD.90.112009
  [arXiv:1408.6457 [hep-ex]].
	
\bibitem{bk} 
  E.~Braaten and M.~Kusunoki,
  Phys.\ Rev.\ D {\bf 71}, 074005 (2005)
  doi:10.1103/PhysRevD.71.074005
  [hep-ph/0412268].
	
\bibitem{panda} 
  D.~Bettoni,
  eConf C {\bf 070805}, 39 (2007)
  [arXiv:0710.5664 [hep-ex]].
	
\bibitem{gottfried} 
  K.~Gottfried,
  Phys.\ Rev.\ Lett.\  {\bf 40}, 598 (1978).
  doi:10.1103/PhysRevLett.40.598
	
\bibitem{mv79} 
  M.~B.~Voloshin,
  Nucl.\ Phys.\ B {\bf 154}, 365 (1979).
  doi:10.1016/0550-3213(79)90037-3

\end{thebibliography}
\end{document}